\documentstyle[11pt,newpasp,twoside,epsfig,subfigure]{article}
\def\edcomment#1{\iffalse\marginpar{\raggedright\sl#1\/}\else\relax\fi}
\begin{document}
\title{Problems encountered in the Hipparcos variable stars analysis}
 \author{Laurent Eyer}
\affil{Instituut voor Sterrenkunde,
             Katholieke Universiteit Leuven,
             Celestijnenlaan 200 B, B-3001 Leuven,
             Belgi\"e}
\author{Michel Grenon}
\affil{Observatoire de Gen\`eve,
             CH-1290 Sauverny,
             Suisse}
\begin{abstract}
Among the 17 volumes of results from the Hipparcos space mission,
two are dedicated to variable stars. These two volumes arose
from the work of two groups, one at the Geneva Observatory
and one at RGO (Royal Greenwich Observatory), on the 13 million
photometric measurements produced by the satellite for
118~204 stars.

The analysis of photometric time series permitted us to identify
several instrumental and mathematical problems: overestimation
of the precision, offsets in the zero-points depending on the field 
of view, mispointing effects, image superpositions, trends in the 
magnitudes, binarity effects (spurious periods and amplitudes)
and time-sampling effects. In this article we summarize some of the 
problems encountered by the Geneva group.
\end{abstract}

\section{Introduction}
The variable star analysis of the Hipparcos photometric data was 
an iterative process in interaction with the FAST and NDAC data
reduction consortia. We started with data extracts from FAST, then
from NDAC and finally worked on the whole merged data set, that is 
on the 118~204 time series.

The result of variable star analysis is a beautiful by-product of 
the mission and it was clear that it had to be published at the same
time as the astrometric results (ESA 1997). The time available for the photometric 
analysis was then short in order to match the deadlines. The teams were
put under strong pressure. The approach was then to produce a robust 
analysis restricting the analysis to statistically well-confirmed 
variables, leaving suspected variables and ambiguous cases for further
analysis.

Because several instrumental problems were identified and solved, the 
variable star study definitely improved the overall quality of the 
Hipparcos photometry available now on the CD-ROMs.

\section{Hipparcos main-mission photometry}
Although the telescope diameter is small (29\,cm), Hipparcos
achieved a high photometric precision in the wide $H\!p$ band (335 to 
895~nm), thanks to the chosen time allocation strategy and to the 
frequent on-orbit photometric calibrations, making use of a large set 
of standard stars. The time allocated for a star observation was adapted 
to its magnitude in order to homogenize the astrometric precision. The 
photometric reduction was made in time slices of 10 hours, called 
reduced great circles (RGC). During that time interval, the satellite 
scanned a closed strip in the sky, measuring about 2600 stars, among 
them 600 standard stars.

The FAST and NDAC consortia independently reduced the photometry. They
had to map the time evolution of the spatial and chromatic response of
the detection chains for both fields of view (FOV), the preceding and
the following.  In Fig.~1, a zero-point problem between FOV magnitude
scales is shown,  before and after its correction.

\begin{figure}
\centering%
  \mbox{
%     \subfigure{\psfig{file=leyer1a.eps,height=70mm,width=70mm}}
%     \subfigure{\psfig{file=leyer1b.eps,height=70mm,width=70mm}}
     \subfigure{\psfig{file=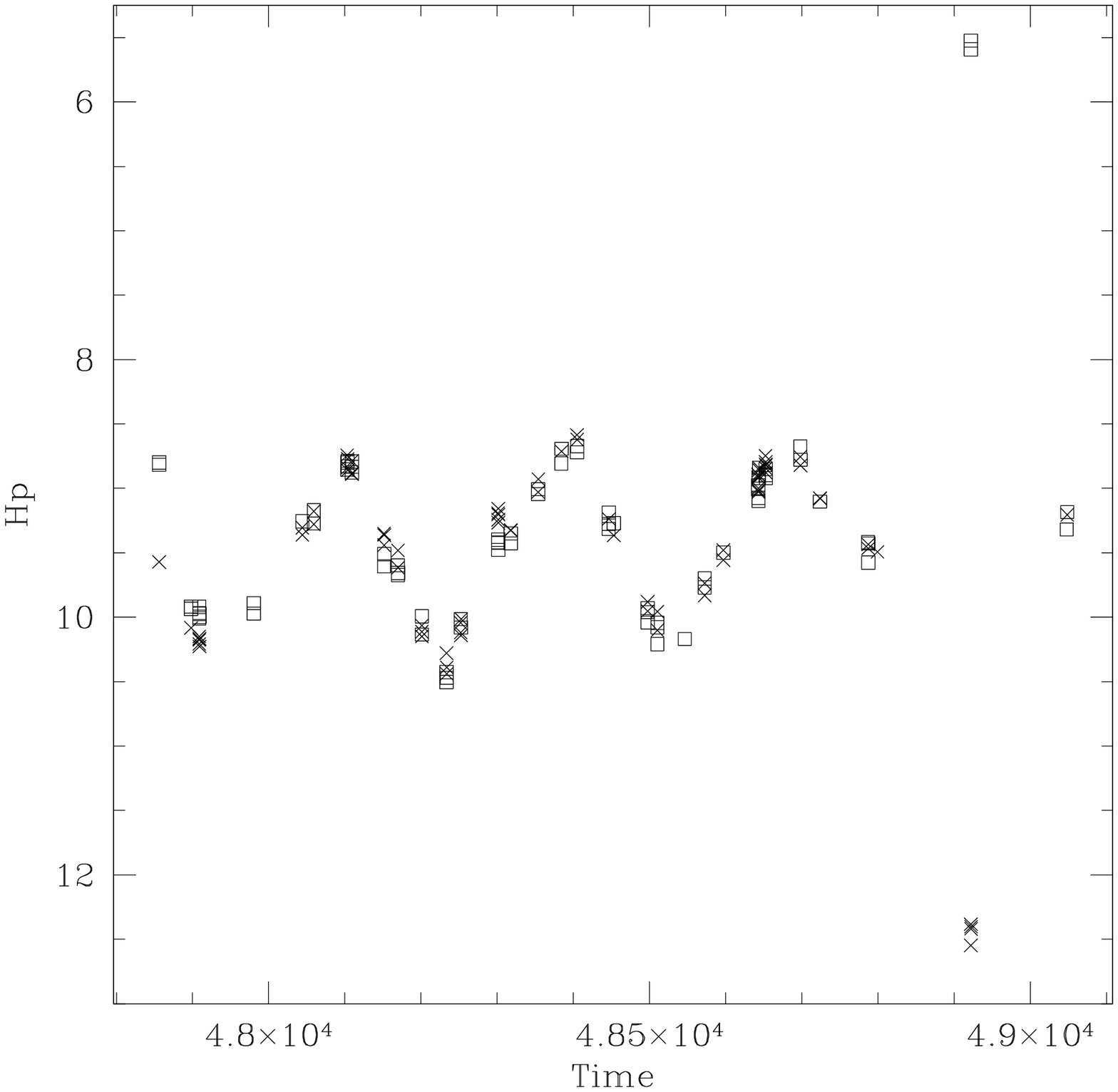,height=66mm,width=66mm}}
     \subfigure{\psfig{file=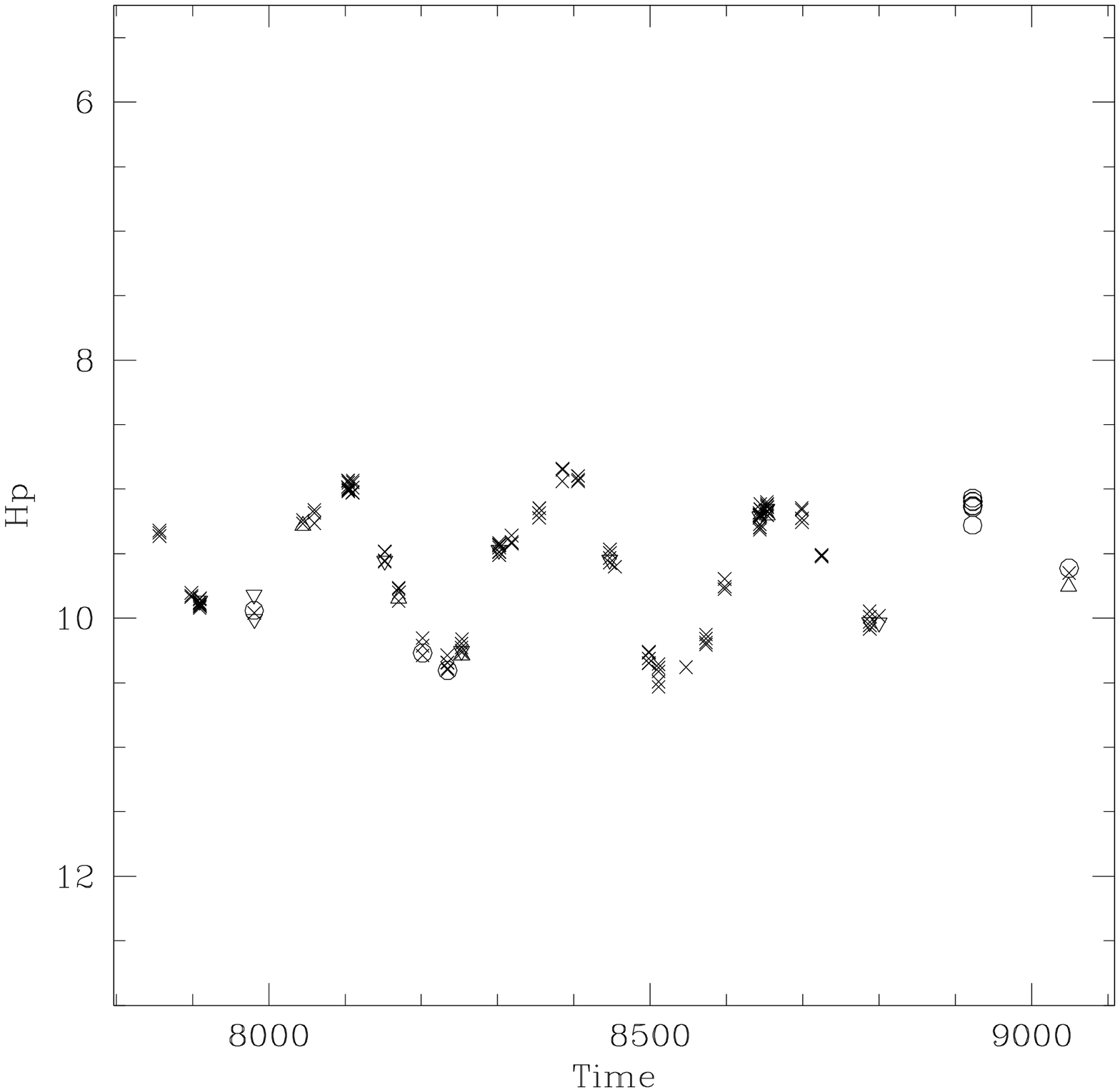,height=66mm,width=66mm}}
   }
  \caption{HIP~96647. Left: magnitudes before the
 data merging, open squares are for the following FOV, crosses for the
 preceding FOV. Right: the corrected final solution.}
\end{figure}

The light of the star was modulated by a grid for astrometric
purposes. The transmitted signal was modeled by a Fourier
series of 5 parameters. From this model two estimates of the
intensity were done: one, measuring the integrated signal, the
``DC mode'', was robust to the duplicity but more dependent on
the background, and the second, measuring the amplitude of the
modulation, the ``AC mode'', was sensitive to the duplicity but
not to the background (cf. van Leeuwen et al. 1997). These two
estimates and their accuracies are given in the Epoch
Photometry Annex (CD-ROM 2) and in the Epoch Photometry Annex
Extension (CD-ROM 3).

\section{Noise and magnitude}
The  precision of the magnitude is a function of the magnitude itself. 
In addition, for a star of constant magnitude, the errors may be 
variable. The data are then heteroscedastic. The correlation between 
the error and the magnitude, may generate some problems. For instance, 
the weighted mean cannot be used to estimate the central value of the 
magnitude distribution, if the amplitude is large as for the Miras. 
The global loss of precision as the satellite ages (Eyer \& Grenon 
2000) also needs to be considered. The usual period search algorithms are 
also sensitive to the inhomogeneity of the data.

\subsection{Quoted transit errors}
During a single transit, a star was measured 9 times on average,
the transit error $\sigma_{H\!p}$ was derived in a first 
approximation from the spread of these measurements. However,
the estimation did not include offsets which might have affected
a whole transit, e.g., due to a mispointing or to a superposition of 
a star from the other FOV. In our first analysis, an empirical law 
was determined to correct the transit error underestimation, 
otherwise the number of candidate variable stars did not appear credible.

During the phase of data merging, ad-hoc corrections were computed
(Evans 1995). The errors were studied with different methods,
comparing first the ``average'' error estimated from the $\sigma_{H\!p}$
with the dispersion of the measurements on $H\!p$. Another study by 
Eyer \& Genton (1999) was made on the quoted errors using variograms;
it showed a good general agreement with the Evans results, with some mild 
underestimations for faint magnitudes and some mild overestimations 
for the bright magnitudes in Evans' approach.

\subsection{Time sampling}
The time sampling was determined by the satellite rotation speed and 
by the scanning law which were optimized to reach the most uniform 
astrometric precision over the whole celestial sphere. The total number
of transits per star is a function mainly of the ecliptic latitude.
The time intervals between successive transits are 20-108-20-etc\ldots 
minutes. The transits form groups which are separated by about one month, 
but the number of consecutive measurements as well as the time separation 
between groups of transits can vary strongly from one star to another.

\subsection{Chromatic aging}
The irradiation by cosmic particles reduced the optical transmission
with time. This aging was chromatic; it was worse than expected
because the satellite had to cross the two van Allen Belts twice per
orbit. Furthermore, the satellite was operational during a maximum of 
solar activity. For instance, the magnitude loss over 3.3 years was 0.8 
mag for the bluest stars and only 0.15 mag for the reddest.

The aging of the image dissector tubes were not uniform and distinct
for both FOVs, therefore the aging corrections had to be calibrated
as functions of the star location on the grid for each FOV.

\subsubsection{Magnitude trends:}
An odd effect of the chromatic aging was the production of magnitude 
trends in the $H\!p$ time series. As the transmission loss was colour 
dependent, the magnitude correction had to be a function of the star 
colour. A colour index, monotonically growing with the effective
wavelength of $H\!p$ band, had to be evaluated from heterogeneous sources.
The precision of the equivalent $V-I$ was highly variable. For stars with
``bad'' $V-I$ colour, the magnitude correction was erroneous and produced
a trend. A colour bluer than true generates a spurious increase of the 
luminosity with the time. An example of a trend is given in
Fig.~2.
\begin{figure}
   \centering%
   \mbox{\psfig{file=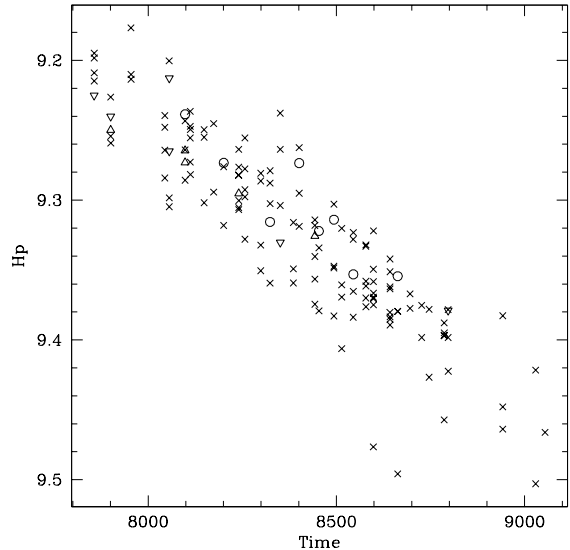,height=80mm,width=80mm}}
   \caption{HIP~29862, an example of trend induced by aging}
\end{figure}
\subsubsection{Selection of trends:}

Stars like Be stars may also show quasi-linear trends over the mission
duration. The identification of spurious trends was iterative.  LPVs,
showing Gaussian residuals when modeled with a trend on top of their
semi-periodic light-curve, were sorted first. But there was much more
diversity in the data showing trends, true or spurious. So we used an
Abbe test (Eyer \& Grenon 2000) for a global detection. Stars with an
Abbe test close to 1, or with a large trend or with very long periods,
were flagged. Stars with possible envelopes were not retained. After
visual inspection of the time series by Grenon, the number of stars
selected by these different procedures was 2412.

\subsubsection{Correction of the star colour:}
The amplitude of the magnitude drift was used to correct the star colour.
Indeed, if a time series shows a trend $\alpha$ which may be imputed to
an incorrect initial colour, there is a possibility to recover the true 
star colour by the relation:
        $$(V-I)_{new} = (V-I)_{old} - 14290 \, \alpha $$
where $\alpha$ is expressed in magnitude per day.

Every selected case was investigated to decide whether the trend could 
be a consequence of an incorrect colour; 965 $V-I$ indices were corrected
this way with certainty and the origin of the errors on the colours was
traced back.

\section{Outlying values}
When studying variable stars, outlying values and anomalous data
distributions are of great interest. Namely, it is important to
distinguish outliers of instrumental origin from those due to stellar
physical phenomena. Some stars show luminosity changes on very short
time scales.  For Algol eclipsing binaries, the duration of the eclipse
is short with respect to their period. With non-continuous time sampling,
eclipses may appear as low luminosity points. UV Ceti stars show strong
bursts in the U band on very short time scales. However, because of the
width of the $H\!p$ band, the photospheric flux in the redder part of
the band largely dominates that of the burst, with the result that no
burst was detected with certainty in the M dwarfs.

In Fig.~3 we present two cases of outlying values of instrumental origin.
\begin{figure}
\centering%
  \mbox{
     \subfigure{\psfig{file=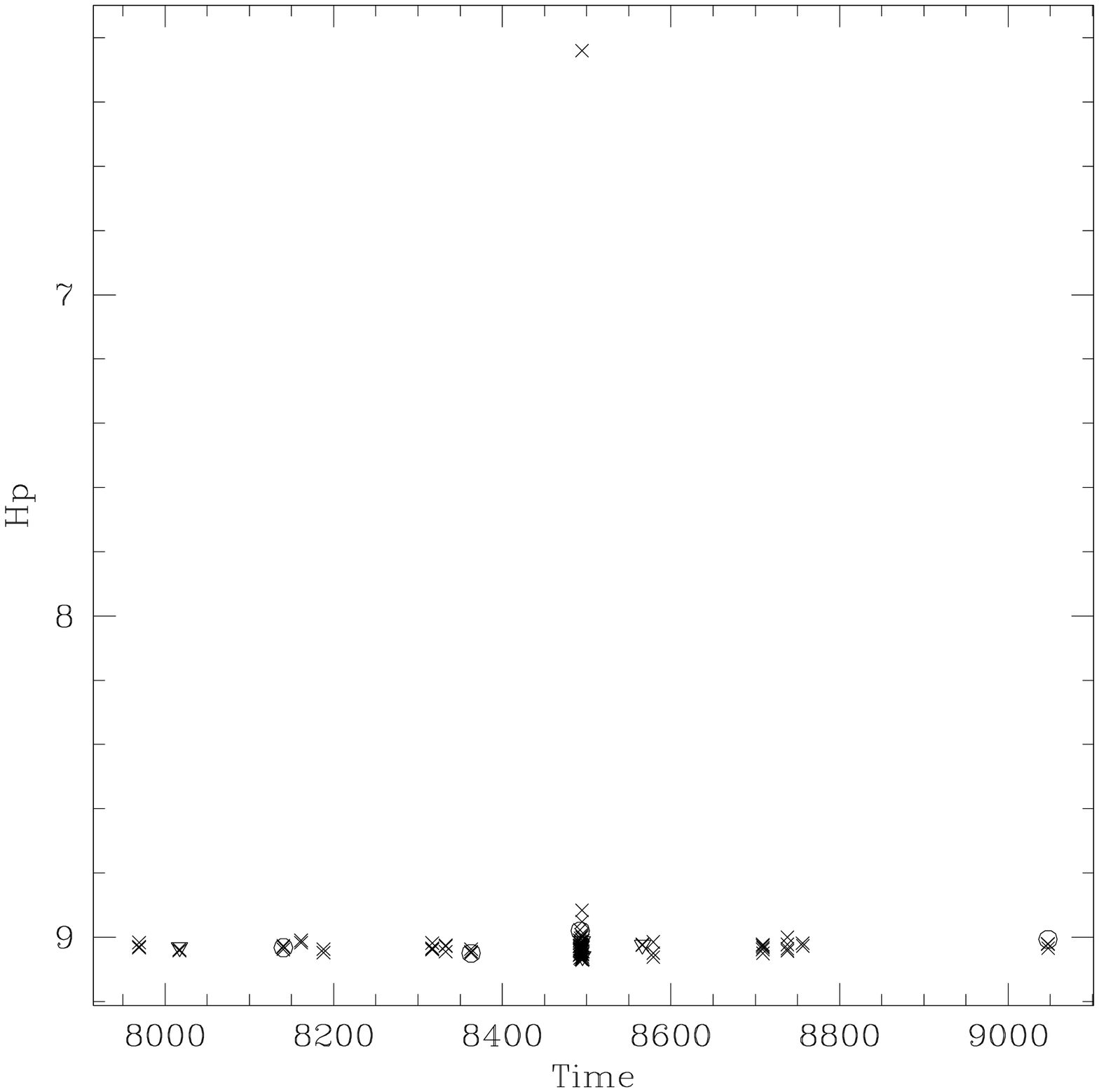,height=66mm,width=66mm}}
     \subfigure{\psfig{file=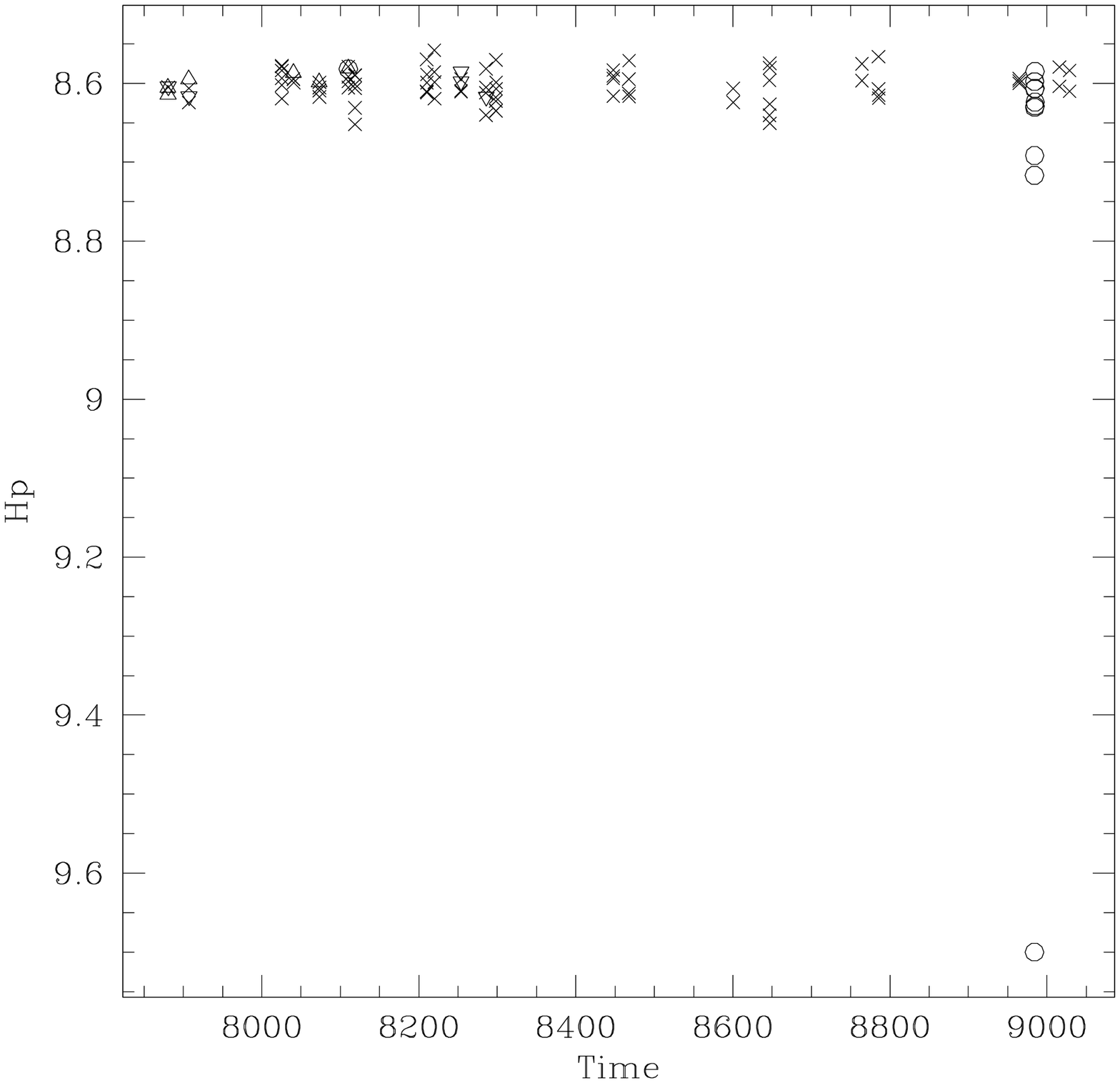,height=66mm,width=66mm}}
   }
  \caption{
    Two examples of series with outlying values. Left: HIP~35527 the case
    of a light pollution from the other FOV, where Sirius is the perturbing 
    star (this case is not flagged); Right: HIP~57437 an example of
    mispointing effect with low values correctly flagged (open circles).}
\end{figure}

\subsection{Instrumental outliers: Mispointing effect}
The pointing precision of the satellite was normally better than 1 arcsec.
However after Earth or Moon eclipses and especially near the end of the 
mission when most gyroscopes were faulty, the problems of mispointing were
more acute. The radius of the photocathode was 15 arcsec, with a lower 
sensitivity towards the edge. An inaccurate pointing was inducing a loss of
counted photons, leading to dimmer points in the time series.

The problems with extended objects were even worse, depending on their sizes.
A similar situation happened with visual double systems when the separation 
was around 10 arcsec. In this case the target was either the primary or the 
photocenter of the system. From time to time the companion was on the edge 
or out of the FOV, diminishing the amount of collected light. Even when the 
two components were measured alternatively, the not-measured star might have
sometimes entered in the FOV producing a luminosity excess mimicking a burst.

\subsection{Instrumental outliers: Light pollution}

A neighbour star could contaminate the observed star, although most of the 
identified cases were rejected during the Input Catalogue compilation.
The perturbing star was possibly a real neighbour or, more often, a star 
belonging to the other FOV. Several configurations are possible:
\begin{itemize}
  \item A star from the other FOV was added to the measured star. That is 
        called a superposition effect. Stars in the Galactic plane were 
        more often perturbed because of the higher star density.
  \item The perturbing star was very bright and caused scattered light in 
        the detection chain. This veiling glare could be felt even if the 
        disturbing star was further than 15 arcsec. 
  \item When the separation was greater that 15 arcsec, the two effects
        of pollution and mispointing could produce high values of fluxes
        of the dim component.
\end{itemize}

Fig.~4 shows the correlation between the asymmetry of 
the time series for double systems and the angular separation $\rho$. The 
asymmetry is positive for dimmer outlying values, and negative for pollution 
by the primary (bright outlying values).
\begin{figure}
   \centering%
   \mbox{\psfig{file=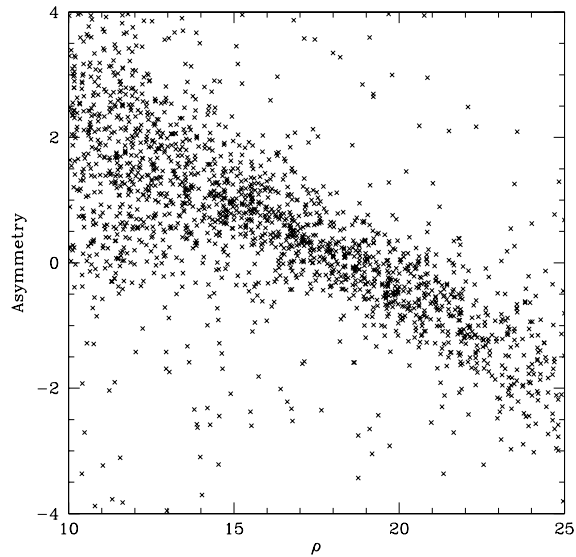,height=80mm,width=80mm}}
   \caption{Light pollution and mispointing effects
for double stars as a function of the angular separation $\rho$.}
\end{figure}

\subsection{Selection of outliers}
The problems caused by the outliers were very acute at the beginning of 
the analysis; we had then to take drastic measures before searching for 
periods and amplitudes. We removed:
\begin{itemize}
  \item all measurements with a non-null flag.
  \item the end of the mission, if the dispersion of the data was smaller 
        than 0.3 before the day 8883. (the data for large amplitude 
        variables were kept up to the end of the mission).
  \item high luminosity values if there was a magnitude jump in consecutive 
        transits.
  \item bad-quality measurements with transit errors higher than 
        $\epsilon(DC) = 0.0005*10^{0.167\: H\!p_{\tiny DC}}+0.0014.$
  \item temporarily one or two outliers to check their impact on the 
        result of an analysis based on truncated time series.
\end{itemize}
This removal represents a reduction by 6\% of the number of measurements
with non-zero flag.

Suspect transits from the analysis of outliers were transmitted to the 
reduction consortia, who flagged them according to the origin of the
disturbance.

\section{Alias and spurious periods}

The spectral window produces spurious periods when it is convoluted 
with the true spectrum. As a result, spurious periods around 0.09 d
were frequently found for long periods or irregular variables as 
well as for stars showing magnitude trends. Periods of about 5 d for
SR variables turned out to be nearly all spurious. With the Hipparcos 
time sampling the spectral window changes from one star to another 
and the alias effect had to be studied on a per star basis. 

A 58 day periodicity was found in many time series when applying the 
period search algorithm. This period corresponds to the time interval 
between consecutive measurements of double systems under the same 
angle with respect to the modulating grid (the modulated signal is 
higher when the components are parallel to the grid).

\section{Advice about the use of the data}
We want to stress that caution should be taken in handling the epoch
photometry, especially when the signal to noise ratio or when the 
number of retained measurements are small. The effects of multiperiodicity 
are generally very tricky. The spectral window should be investigated 
in detail and periods near sampling frequencies should be taken with 
care, in particular in the range 5 to 20 d where the Hipparcos photometry
has the weakest detection capability.

The selection of photometric data can be made according to their flags,
the estimates of the transit errors and the background intensities.
In case of doubt about outlying data, a look to the magnitude difference
$AC-DC$ will reveal problems related to duplicity and image superpositions
since the amplitude of the modulated signal is reduced in the case of
misaligned sources with respect to the grid orientation. The contents of
the opposite FOV can be investigated thanks to their published positions.
It is suggested to correct rather than to eliminate data since a loss
of information might twist statistics. For an example of a successful
selection procedure applied to the data of HIP~115510, see Lampens et
al. (1999).

\section{Conclusion}
Performing accuracy photometry in space is not free from problems. The
same is true for the data analysis. Once the origin of the encountered 
problems is identified, it is possible to cope with them and determine
precisely the domains of validity of the algorithms for search of periods
and amplitudes. Globally the ratio quantity-quality generated by this 
mission for the study of variability has no equivalent up to now.

\end{document}